\documentclass[%
 reprint,
showpacs,preprintnumbers,
 amsmath,amssymb,
 aps,
 prc,
]{revtex4-1}

\usepackage{graphicx}
\usepackage{dcolumn}
\usepackage{bm}

\begin{document}

\preprint{APS/123-QED}

\title{Beam energy dependence and updated test of the Trojan horse nucleus invariance \\ via the d(d,p)t measurement at low energies \\ 
 \thanks{ }}

\author{Chengbo Li}
\email{lichengbo2008@163.com}
\thanks{Supported by Beijing Natural Science Foundation (1122017) and National Natural Science Foundation of China (11075218, 10575132) }
\affiliation{Beijing Radiation Center, Beijing 100875, China}
\affiliation{Key Laboratory of Beam Technology and Material Modification of Ministry of Education,College of Nuclear Science and Technology, Beijing Normal University, Beijing 100875, China}

\author{Qungang Wen}
\affiliation{School of Physics and Materials Science, Anhui University,  Hefei 230601, China}

\author{Yuanyong Fu}
\author{Jing Zhou}
\author{Shuhua Zhou}
\author{Qiuying Meng}
\affiliation{Department of Nuclear Physics, China Institute of Atomic Energy, Beijing 102413, China}

 \author{C. Spitaleri}
 \author{A. Tumino}
 \author{R. G. Pizzone}
 \author{L. Lamia}
 \affiliation{Laboratori Nazionali del Sud, INFN, Catania, Italy}

%
%
%

\date{\today}

\begin{abstract}

The $\mathrm{^2H}(d,p)\mathrm{^3H}$ bare nucleus astrophysical S(E) factor has been measured indirectly at energies from about 500 keV down to several keV by means of the Trojan-horse method applied to the quasi-free process $\mathrm{^2H({}^6Li},pt)\mathrm{^4He}$ induced at the lithium beam energy of 11 and 9.5 MeV, which makes the virtual binary process incident energy $\mathrm {E}_{dd}^{qf}$ go much closer to the zero-quasi-free-energy point than that in the previous similar experiment. 
The obtained results are compared with direct data as well as with previous indirect investigation of the same binary reactions.
It shows that the precision of S(E) data in low energy range extracted via the same Trojan horse nucleus ($\mathrm{^6Li}=(d \oplus \alpha)$ ) becomes better when the incident energy decreases from high value down to the zero-quasi-free-energy point.  
The very good agreement between data extracted from different Trojan horse nucleus ($\mathrm{^6Li}=(d \oplus \alpha)$ vs. $\mathrm{^3He}=(d \oplus p)$) gives a strong updated test for the independence of the binary indirect cross section on the chosen Trojan horse nucleus at low energies.

\end{abstract}



                              
\maketitle


\section{Introduction}
In the last decades, the improvements in the field of astrophysics observation and models have triggered nuclear reaction measurements at the astrophysical energies \cite{wall97}. Because of their crucial role in understanding the first phases of the universe history and the subsequent stellar evolution, cross sections at the Gamow energy $E_G$ should be known with an accurate value. 
For example, The important $d+d$ nuclear reaction we are interested in this work takes place at a ultra low energy of about several keV to 300 keV in astrophysical environment: the region of interest ranges from 50 to 300 keV as for the Standard Big Bang Nucleosynthesis, and it is only from a few to 20 keV for stellar evolution processes. 

For charged particle induced reactions the Coulomb barrier $E_C$, usually of the order of MeV, is much higher than $E_G$ (typically smaller than few hundred keV), thus implying that as the energy is lowered the reactions take place via tunneling with an exponential decrease of the cross section. 
The direct measurement of nuclear reactions induced by charged particles at astrophysical energies has many experimental difficulties.
The difficulties mainly come from the presence of the strong Coulomb suppression and the electron screening effect which would lead to a screened cross section larger than the bare nucleus one at ultra-low energies \cite{thmrev2014}.

To overcome these difficulties, many efforts were devoted to the development and application of indirect methods in nuclear astrophysics\cite{thmrev2014}. Among the most-used indirect methods, the Trojan horse method (THM) \cite{gbaur1986, stypel2003} played an important role, which has been applied to several reactions at the energies relevant for astrophysics in the past decade \cite{thmrev2014}. 

Many tests have been made to fully explore the potential of the method and extend as much as possible its applications \cite{cspitaleri2003,thmrev2011,thmrev2014,thmdist2009,ths,thsli}: the target-projectile breakup invariance, the spectator invariance, and the possible use of virtual neutron beams. 
Such studies are necessary, because the Trojan horse method has become one of the major tools for the investigation of reactions of astrophysical interest\cite{thmrev2014,thmrev2011}. 

In recent works \cite{thsli} the spectator invariance was extensively examined for the $\mathrm{^6Li({}^6Li},\alpha \alpha)\mathrm{^4He}$ and the $\mathrm{^6Li({}^3He},\alpha \alpha)\mathrm{^1H}$ case as well as the $\mathrm{^7Li}(d,\alpha \alpha)n$ and $\mathrm{^7Li({}^3He},\alpha \alpha)\mathrm{^2H}$ reactions, thus comparing results arising from $\mathrm {^6Li}=(d \oplus \alpha)$ and $\mathrm {^3He}=(d \oplus p)$ breakup to provide virtual $d$, and $d=(p \oplus n)$ and $\mathrm {^3He}=(p \oplus d)$ breakup to provide virtual $p$, respectively .
Agreement between the sets of data was found below and above the Coulomb barrier. This suggests that $\rm {^3He}$ is a good Trojan horse nucleus in spite of its quite high $\mathrm {^3He}\rightarrow d + p$ breakup energy (5.49 MeV) and that the THM cross section does not depend on the chosen Trojan horse nucleus, at least for the processes mentioned above.

The Trojan Horse Method has also been applied to the indirect study of the $d+d$ reactions using $\mathrm {^3He}=(d \oplus p)$  and $\mathrm {^6Li}=(d \oplus \alpha)$ breakup \cite{thmdd2013,thmdd2014} to test the Trojan horse nucleus invariance, but the $\rm ^6Li$ breakup data give much less points and larger errors than that in the case of $\rm ^3He$  breakup. The Trojan horse nucleus invariance can not be well confirmed for this case with such big errors.

In this paper, we report on a new investigation of the $\mathrm{^2H}(d,p)\rm{^3H}$ reaction by means of the THM applied to the $\mathrm {^2H({}^6Li},pt)\rm{}^4He$ quasi-free process with the beam energy of 11 MeV and 9.5 MeV, which makes the virtual binary process incident energy $\mathrm {E}_{dd}^{qf}$ go much closer to the zero quasi-free energy point than that in \cite{thmdd2013} with the beam energy of 14 MeV. 
We will study the dependence of the S(E) data precision on the incident beam energy with same Trojan horse nucleus.
At last, an updated test will be given to the Trojan horse nucleus invariance for the case of $d(d,p)t$  reaction at low energies (Figure \ref{figTHM} (b)(c)).

\section{Trojan Horse Method}
The Coulomb barrier and electron screening cause difficulties in directly measuring nuclear reaction cross sections of charged particles at astrophysical energies. 
To overcome these difficulties, the THM \cite{gbaur1986, stypel2003} has been introduced as a powerful indirect tool in experimental nuclear astrophysics \cite{cspitaleri2003, thmall2013, thmrev2011, thmrev2014, thmdist2009, thmplb1, thmplb2, thmdd2013, thmdd2014, ths, thsli, cheng2005, sromano2006, wen2008, wen2011, wen2016, thmg4}. 
The THM provides a valid alternative approach to measure unscreened low-energy cross sections of charged particle reactions. It can also be used to retrieve information on the electron screening potential when ultra-low energy direct measurements are available.

The basic assumptions of the THM theory have been discussed extensively in \cite{ stypel2003, cspitaleri2003, thmall2013,thmrev2011,thmrev2014},  and the detailed theoretical derivation of the formalism employed can be found in \cite{stypel2003} .

\begin{figure*}
\begin{center}
\includegraphics[width = 0.75\textwidth]{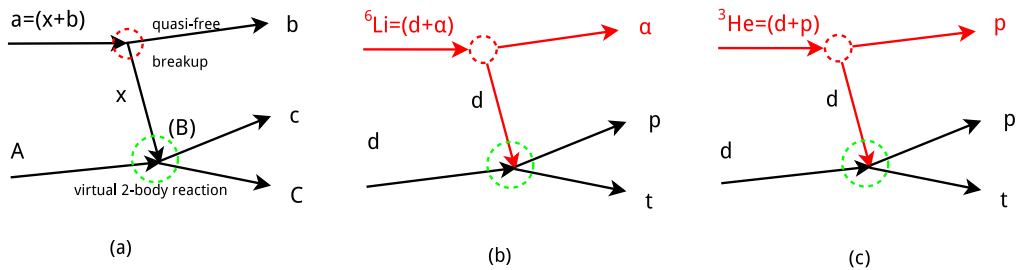}\\
\caption{(Color online) The schematic representation of Trojan-horse method. (a) Normal, (b) TH: $\rm ^6Li$, (c) TH: $\rm ^3He$ }
\label{figTHM} 
\end{center}
\end{figure*}

A schematic representation of the process underlying the THM is shown in Figure \ref{figTHM}.
The method is based on the quasi-free (QF) reaction mechanism, which allows one to derive indirectly the cross section of a two-body reaction Eq.(\ref{eq:two_body})
\begin{equation}\label{eq:two_body}
    A+x \rightarrow C+c
\end{equation}
from the measurement of a suitable three-body process under the quasi-free kinematic conditions: 
\begin{equation}\label{eq:three_body}
    A+a \rightarrow C+c+b
\end{equation}
where the nucleus $a$ is considered to be dominantly composed of clusters $x$ and $b$ ( $a=(x \oplus b) $).

After the breakup of nucleus $a$ due to the interaction with nucleus $A$, the two-body reaction  (Eq.(\ref{eq:two_body})) occurs only between nucleus $A$ and the transferred particle $x$ whereas the other cluster $b$ behaves as a spectator to the virtual  two-body reaction during the quasi-free process. 
The energy in the entrance channel $\rm E_{Aa}$ is chosen above the height of the Coulomb barrier $\rm E_{Aa}^{C.B.}$ , so as to avoid the reduction in cross section. 

At the same time, the effective energy $\rm E_{Ax}$ of the reaction between $A$ and $x$ can be relatively small, mainly because the energy $\rm E_{Aa}$ is partially used to overcome the binding energy $\varepsilon_a$ of $x$ inside $a$, even if particle $x$ is almost at rest the extra-energy is compensated by the binding energy of $a$  (Eq.(\ref{eq:Eqf1})), and the Fermi motion of $x$ inside $a$, $\rm E_{xb}$, is used to span the region of interest around $\rm E_{Ax}^{qf}$:
\begin{equation}\label{eq:Eqf1}
        E_{Ax}^{qf}=E_{Aa}\left(1-\frac{\mu_{Aa}}{\mu_{Bb}}\frac{\mu_{bx}^{2}}{m_x^2}\right)-\varepsilon_{a}
\end{equation}
\begin{equation}\label{eq:Eqf2}
        E_{Ax}=E_{Ax}^{qf}\pm E_{xb}
\end{equation}

Since the transferred particle $x$ is hidden inside the nucleus $a$ (so called Trojan-horse nucleus), it can be brought into the nuclear interaction region to induce the two-body reaction $A+x$, which is free of Coulomb suppression and, at the same time, not affected by electron screening effects.

Thus, the two-body cross section of interesting can be extracted from the measured quasi free three-body reaction inverting the following relation:
\begin{equation}\label{eq:sec3all}
\frac{d^3\sigma}{dE_{Cc}d\Omega_{Bb}d\Omega_{Cc}} = K_F \cdot |W|^2  \cdot  \frac{d\sigma}{d\Omega}^{TH} 
\end{equation}
where $\rm K_F$ is the kinematical factor,  $\rm |W|^2$ is the momentum distribution of the spectator $b$ inside the Trojan-horse nucleus $a$,   
and $\rm {d\sigma}/{d\Omega}^{TH}$ is the half-off-energy-shell (HOES) cross section of the two-body reaction $\rm A+x\rightarrow C+c$:
\begin{equation}\label{eq:sec2all}
\frac{d\sigma}{d\Omega}^{TH} = \sum_l C_l \cdot P_l  \cdot \frac{d\sigma_l}{d\Omega}({Ax\rightarrow Cc)}
\end{equation}
where $\rm \frac{d\sigma_l}{d\Omega}({Ax\rightarrow Cc)}$ is the real on-energy-shell cross section of the two-body reaction $\rm A+x\rightarrow C+c$ for the $l$ partial wave,  $\rm P_l$ is the penetration function caused by the Coulomb wave function, and $\rm C_l$ is the scaling factor.

The relevant two-body reaction cross section can be extracted from the measured three-body cross section with cut of selecting the quasi-free events from the three-body reaction.
Then, the S(E) factor can be determined from the definition below, where the Sommerfeld parameter is $\rm \eta=Z_1 Z_2 e^2 / (\hbar v)$.
\begin{equation}\label{eq:SE}
    S(E)= \sigma (E) E \exp(2\pi\eta)
\end{equation}

The lack of screening effects in the THM $\rm S_{bare}(E)$ factors gives the possibility to return the screening potential $\rm U_e$ from comparison with direct data using the following screening function with $\rm U_e$ as free parameter. 
\begin{equation}\label{eq:Ue}
    f_{lab}(E)=\sigma_s(E)/\sigma_b(E)\simeq \exp(\pi \eta U_e /E)
\end{equation}

\section{Experiment}

\begin{figure}
\begin{center}
\includegraphics[width = 0.45\textwidth]{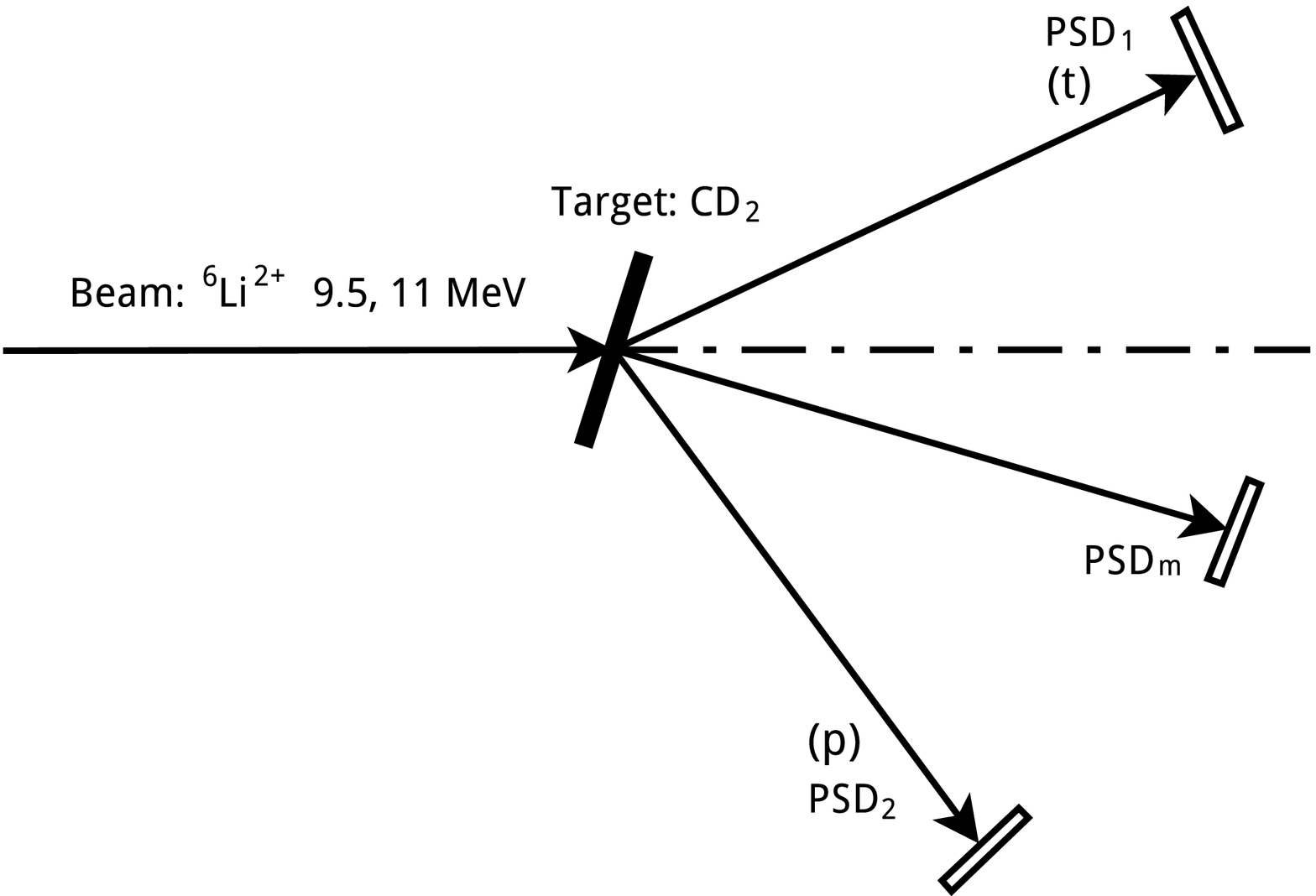}\\
\caption{Experiment setup of the $\rm {}^2H({}^6Li,pt){}^4He$ reaction measurement.}
\label{figSetup} 
\end{center}
\end{figure}

The measurement of the $\rm {}^2H({}^6Li,pt){}^4He$ reaction was performed at the Beijing National Tandem Accelerator Laboratory at China Institute of Atomic Energy. 
The experimental setup was installed in the nuclear reaction chamber at the R60 beam line terminal as shown in Figure \ref{figSetup}. 
The $\rm ^{6}Li^{2+}$ beam at 9.5 MeV \cite{thmdd2015} and 11 MeV provided by the HI-13 tandem accelerator was used to bombard a deuterated polyethylene target $\rm CD_{2}$. 
The thickness of the target is about $\rm 160 \mu g/cm^{2}$. In order to reduce the angle uncertainty coming from the large beam spot, a strip target with 1 mm width was used.

A position sensitive detector $\rm PSD_1$ was placed at $40^{\circ} \pm 5^{\circ}$ to the beam line direction and about 238 mm from the target to detect the outgoing particle triton ($t$), 
and another detector $\rm PSD_2$ was used at $78^{\circ} \pm 5^{\circ}$  in the other side of the beam line at 245 mm  from the target to detect the outgoing particle proton ($p$).
The arrangement of the experimental setup was modelled in Monte Carlo simulation in order to cover a region of quasi-free angle pairs.
A $\rm PSD_m$ was placed at $32^{\circ} \pm 5^{\circ}$ opposite to $\rm PSD_1$ as a monitor.
The energy resolution of the PSDs is about 0.6\%-0.8\% for 5.48 MeV $\alpha$ source. 

It should be mentioned that no $\rm \Delta E$ detector was mounted before PSDs, that was, no particle identification was performed in the experiment. 
It was beneficial to improve the energy and angular resolution without $\rm \Delta E$ detectors, because it would lead to additional straggling of energies and angles when particles passing through these detectors which are mounted in \cite{thmdd2013}. 
While it was easy to select events for this reaction without particle identification from the kinematical locus in data analysis with the help of simulation. 

The trigger for the event acquisition was given by coincidence of signals by $\rm Gate=PSD_1\times(PSD_2+PSD_m)$.
Energy and position signals for the detected particles were processed by standard electronics and sent to the acquisition system MIDAS for on-line monitoring and data storage for off-line analysis.

The position and energy calibration of the detectors  were performed using elastic scatterings on different targets ($\rm ^{197}Au$, $\rm ^{12}C$, and $\rm CD_2$) induced by a proton beam at energies of 6, 7, 8 MeV.  A standard $\alpha$ source of 5.48 MeV was also used. 
In order to perform position calibration, a grid with a number of equally spaced slits was placed in front of each PSD for calibration runs.

\section{Data analysis and results}

After the calibration of the detectors, the energy and momentum of the third undetected particle ($\alpha$)  were reconstructed from  the complete kinematics of the three-body reaction $\mathrm {^6Li} + d \rightarrow t+p+ \alpha$, under the assumption that the first particle is a triton (detected by $\rm PSD_1$)  and the second one is a proton (detected by $\rm PSD_2$).
The detailed data analysis for 9.5 MeV runs can been seen in ref \cite{thmdd2015}, and the method is similar for 11 MeV runs.

\subsection{Selection of the quasi-free three body reaction events}

The first step of data analysis is to select the three-body reaction events of $\rm {}^2H({}^6Li,pt){}^4He$ from all exit channels of the experiment.
Comparing with the Monte Carlo simulation \cite{thmg4}, we can easily select the three body reaction $\rm {}^2H({}^6Li,pt){}^4He$ events from the experimental spectrum of the $\rm E_1- E_2$ kinematic locus by a graphical cut where they are located \cite{thmdd2015}.
It will be used as a basic selection cut in the following data analysis.

\begin{figure}
\begin{center}
\includegraphics[width = 0.45\textwidth]{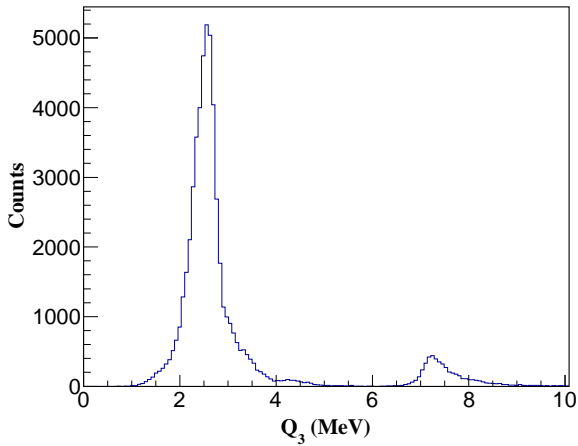}\\
\caption{(Color online) Experimental $\rm Q_3$ value spectrum for 11 MeV runs with the kinematic locus cuts. The peak around 2.5 MeV is in good agreement with the theoretical prediction value of $\rm {}^2H({}^6Li,pt){}^4He$ (Q = 2.558 MeV).}
\label{figQ3} 
\end{center}
\end{figure}

Once selected the three-body reaction events of $\rm ^{2}H(^{6}Li, pt)^{4}He$, the experimental $\rm Q_3 $ value can be extracted for the 11 MeV runs, as reported in Figure \ref{figQ3}. 
There is a peak whose centroid is at about 2.5 MeV (in good agreement with the theoretical prediction, Q = 2.558 MeV). 
It is a clear signature of the good calibration of detectors as well as of the correct identification of the reaction channel. 
Only events inside the 2.5 MeV Q-value peak were considered for the further analysis.

\begin{figure}
\begin{center}
\includegraphics[width = 0.45\textwidth]{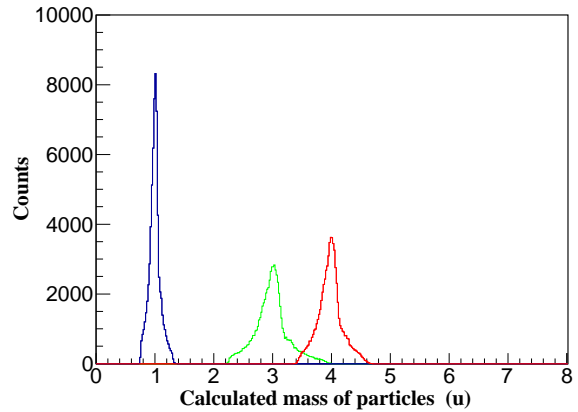}\\
\caption{(Color online) Calculated mass spectrum of the three outgoing particles. Green: $\rm m_1\simeq 3$ ($t$), red: $\rm m_2\simeq 1$ ($p$), black: $\rm m_3\simeq 4$ ($\alpha$).}
\label{figMass} 
\end{center}
\end{figure}

The mass of the three outgoing particles can be calculated by the kinematic relations when the theoretical value $Q_3$=2.558 MeV was given with the cuts  selected as above for the three-body reaction events of $\rm ^{2}H(^{6}Li, pt)^{4}He$. The calculated mass spectrum was shown in Figure \ref{figMass}. The mass number of the first particle $\rm m_1 \simeq 3$, it was agree with triton ($t$), and $\rm m_2 \simeq 1$ ($p$), $\rm m_3 \simeq 4$ ($\alpha$). It gave a good test to the particle identification with kinematics.

The obtained momentum distribution for 11 MeV runs which is similar to \cite{thmdd2015} gives a main test for the presence of the quasi-free mechanism and the possible application of the THM.
For the further analysis, the condition of $\rm |p_{s}|<20 MeV/c$ will be added to the above cuts to select the quasi-free events of the three-body reaction.

\subsection{Results and discussion: S(E) factor }

At last, the energy trend of the S(E) factor can be extracted by means of the standard procedure of the THM after selecting the quasi-free three-body reaction events \cite{thmdd2015}. 

\begin{figure} 
\begin{center}
\includegraphics[width = 0.35\textwidth,angle=-90]{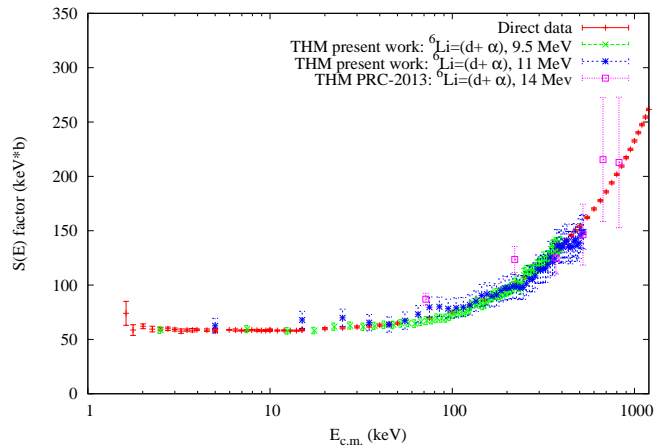}\\
\caption{(Color online) The S(E) factor obtained from THM measurements with same TH ($\rm ^6Li=(d \oplus \alpha)$) and different energy compared with direct data}
\label{figSELi6} 
\end{center}
\end{figure}

\begin{figure} 
\begin{center}
\includegraphics[width = 0.35\textwidth,angle=-90]{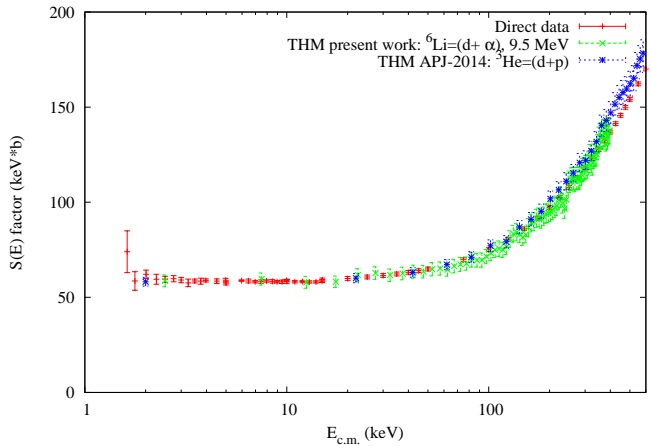}\\
\caption{(Color online) The S(E) factor obtained from different THM measurements with different TH ($\rm ^6Li=(d \oplus \alpha)$ vs. $\rm ^3He=(d \oplus p)$) compared with direct data to test the Trojan horse nucleus invariance. }
\label{figSE} 
\end{center}
\end{figure}

\begin{table*}
\begin{center}
    \caption{\label{tab:THMdata} Comparison of d(d,p)t indirect studies via THM.}
    \begin{ruledtabular}
        \begin{tabular*}{170mm}{c@{\extracolsep{\fill}}ccccc}
            Work & TH & $\rm E_0$ (MeV)  & $\mathrm {E}_{dd}^{qf} $ (keV) & $\rm S_0(E)$  ($\rm keV \cdot b$) & $\rm U_e $ (eV) \\
             \hline
            Our work \cite{thmdd2015}  & $\rm ^6Li=(d \oplus \alpha)$  & 9.5   & 89 & $56.7 \pm 2.0$ & $13.2 \pm 4.3$\\
            Present work  & $\rm ^6Li=(d \oplus \alpha)$  & 11   & 342 & $61.1 \pm 6.8$ & $19.6 \pm 9.2$\\
            PRC-2013 \cite{thmdd2013} & $\rm ^6Li=(d \oplus \alpha)$  & 14  & 866 & $75 \pm 21$ & -\\
            APJ-2014 \cite{thmdd2014} & $\rm ^3He=(d \oplus p)$  & 17  & 178 &  $57.7 \pm 1.8$ & $13.4 \pm 0.6$\\
      \end{tabular*}
    \end{ruledtabular}  
\end{center}
\end{table*}

The results of the bare nucleus astrophysical $\rm S_{bare}(E)$ factor for the $\rm d(d, p)t$ reaction are presented in Figure \ref{figSELi6} (green points for 9.5 MeV runs, and blue points for 11 MeV runs) after normalization with direct data (red points) \cite{thmdd2014, direct}, and they are compared with the data from PRC-2013 \cite{thmdd2013} using the same Trojan horse nucleus $\rm ^6Li=(d \oplus \alpha)$ breakup in a previous THM experiment (pink points). The energy information together with $S_0(E)$ and $U_e$ data are also shown in Table \ref{tab:THMdata}.

It should be pointed out that direct data suffer from the electron screening effect which does not affect the THM results. Thus, the S(E) extracted via THM was called $\rm S_{bare}(E)$. In addition, $\rm U_e$ can be extracted \cite{thmdd2015} by comparing THM data with direct ones.
The normalization was performed in the energy range where the electron screening effect is still neglectable.

It shows that the errors of both 9.5 MeV runs and 11 MeV runs are much smaller than that in PRC-2013 \cite{thmdd2013} using the same Trojan horse at higher beam energy (14 MeV) with $\rm \Delta E$ detectors for particle identification in the experimental setup.

The good results mainly due to the three effective disposition we performed:
First, the strip target with 1 mm width was used to reduce the angle uncertainty coming from the large beam spot.
Second, no $\rm \Delta E$ detector was mounted before PSDs. It was beneficial to improve the energy and angular resolution, which can avoid the additional straggling of energies and angles when particles passing through $\rm \Delta E$ detectors if they were mounted. And the particle identification can be done during the data analysis with the help of kinematics (as shown in Fig.\ref{figMass}).
Finally, the beam energy was reduced to 11 and 9.5 MeV, which lead to the quasi-free binary incident energy $\rm E_{dd}^{qf}$ went closer to the zero energy point than that in PRC-2013 \cite{thmdd2013}, shown in Table \ref{tab:THMdata}.

The data precision of 9.5 MeV runs are better than 11 MeV runs in low energy range, although the data of 11 MeV runs can extend to higher energies from 400 keV to 500 keV than that of 9.5 MeV runs.

In summary, it can be seen from Table \ref{tab:THMdata} and Fig.\ref{figSELi6} that the precision of S(E) data in low energy range extracted via the same Trojan horse nucleus becomes better when the incident energy decreases from high value down to the zero-quasi-free-energy point. That is the main reason of why the PRC-2013 \cite{thmdd2013} data can not give a perfect comparison to APJ-2014 \cite{thmdd2014} data.

The data from the present experiment (green points for 9.5 MeV runs \cite{thmdd2015} ) are also compared with those from APJ-2014 \cite{thmdd2014} (blue points) using different Trojan horse nucleus $\rm ^3He=(d \oplus p)$, shown in Fig \ref{figSE}.
An overall agreement is present among both direct and indirect data sets with different Trojan horse nucleus, within the experimental errors, i.e., data extracted via the THM applied to  $\rm ^6Li$  and $\rm ^3He$ breakup are comparable among themselves. 
That is, the use of a different spectator particle does not influence the THM results.
Thus, the Trojan horse particle invariance, which was already observed in Ref.\cite{ths}, is confirmed in an additional and independent case. 
This work gives a much strong updated test for the independence of the binary indirect cross section on the chosen Trojan horse nucleus at low energies than the previous test in PRC-2013 \cite{thmdd2013}.

\section{Summary}

A new investigation of the $\rm ^2H({}^6Li,pt){}^4He$ reaction measurement to extract information on the astrophysical $\rm S_{bare}(E)$ factor for the $\rm d(d, p)t$ reaction via the THM at lower incident energies (11 MeV and 9.5 MeV) is presented in the present paper. 
The obtained results are compared with direct data as well as with previous indirect investigation of the same binary reactions \cite{thmdd2013,thmdd2014}.

It shows that the precision of S(E) data in low energy range extracted via the same Trojan horse nucleus becomes better when the incident energy decreases from high value down to the zero-quasi-free-energy point. It is the main reason of why the previous PRC-2013 \cite{thmdd2013} experiment can not get a good result comparison to APJ-2014 \cite{thmdd2014} data.

The very good agreement between data extracted from different Trojan horse nucleus ($\mathrm{^6Li}=(d \oplus \alpha)$ vs. $\mathrm{^3He}=(d \oplus p)$ \cite{thmdd2014})  gives a much strong updated test than the previous test \cite{thmdd2013} for the Trojan horse nucleus invariance at very low energies energies.

\begin{acknowledgments}
We thank Dr. LIN Chengjian, Dr. JIA Huiming and other persons in their research group for their kind help during the experimental preparation and measurement. We thank Dr. LI Xia for the help of the acquisition system, and the CIHENP research group for their help during the experiment. 
We also thank the staff of the HI-13 tandem accelerator laboratory for providing experimental beam and targets.

\end{acknowledgments}


\end{document}